\documentclass[11p]{scrartcl}
 \usepackage[top=2.5cm, bottom=2.5cm, left=2.5cm, right=2.5cm]{geometry}

\usepackage[T1]{fontenc}
\usepackage[utf8]{inputenc}
\usepackage{microtype}
\usepackage[english]{babel}

\usepackage{pdfpages}

\usepackage{amsmath}
\usepackage{amsfonts}
\usepackage{amssymb}
\usepackage{mathrsfs}
\usepackage{amstext}
\usepackage{bbold}
\usepackage{mathtools}
\usepackage{physics}
\usepackage{nicefrac}
\usepackage{dsfont}
\usepackage{cleveref}

\usepackage{graphicx}
\usepackage{wrapfig}
\usepackage{multicol}
\usepackage[e]{esvect}
\usepackage[per-mode=reciprocal, exponent-product=\cdot, separate-uncertainty=true, emulate=units]{siunitx}
\usepackage{gensymb}
\usepackage{amsthm}
\usepackage{subfig}
\usepackage{caption}
\usepackage{xcolor}
\usepackage{placeins}

\usepackage{array}
\usepackage{tabularx}
\usepackage{tabulary}
\usepackage{booktabs}
\usepackage{longtable}

\newcommand*{\IC}{\mathds{C}}

\newcommand*{\al}[1]{\mathfrak{#1}}

\newcommand*{\oper}[1]{\hat{#1}}

\newcolumntype{C}{>{$}c<{$}}
\newcolumntype{L}{>{$}l<{$}}
\newcolumntype{R}{>{$}r<{$}}

\newcommand*{\jz}{\oper{j}_z}
\newcommand*{\kp}{\oper{k}_+}
\newcommand*{\kx}{\oper{k}_x}
\newcommand*{\ky}{\oper{k}_y}

\newcommand{\km}{\oper{k}_-}
\newcommand*{\kpm}{\oper{k}_\pm}

\newcommand*{\JZ}{\oper{J}_z}
\newcommand*{\KP}{\oper{K}_+}
\newcommand{\KM}{\oper{K}_-}
\newcommand*{\KPM}{\oper{K}_\pm}
\newcommand*{\KX}{\oper{K}_x}
\newcommand*{\KY}{\oper{K}_y}

\newcommand*{\fac}{!\,}

\newcommand*{\partsum}[1]{\sum_{\substack{{#1}_1,\ldots {#1}_N=0\\\sum_i{#1}_i=#1}}}

\usepackage[normalem]{ulem}

\title{\textsf{{A note on coarse graining and group representations}}\vspace{0.35cm}}

\author{
{Norbert Bodendorfer\footnote{\texttt{norbert.bodendorfer@physik.uni-r.de}}, Fabian Haneder\footnote{\texttt{fabian.haneder@physik.uni-r.de}}}\\
{{Institute for Theoretical Physics, University of Regensburg,}}\\
{{93040 Regensburg, Germany}}
}

\renewcommand{\bf}{\bfseries}
\renewcommand{\tt}{\ttfamily}

\begin{document}

\maketitle

\begin{abstract}
{A coarse graining operation of spatially homogeneous quantum states based on an SU(1,1) Lie group structure has recently been proposed in \cite{BodendorferCoarseGrainingAs} and used in \cite{BWI} to compute an explicit renormalisation group flow in the context of loop quantum cosmology. In this note, we explain the group theoretical origin of this procedure and generalise previous results based on these insights. We also highlight how the group theoretical origin of these techniques implies their immediate generalisation to other Lie groups. 
 }
\end{abstract}

\section{Introduction}

The coarse graining of quantum states and the associated notion of a renormalisation group flow of quantum states and operators is one of the main open questions in loop quantum gravity, see \cite{ThiemannRenormalisationReview, SteinhausCoarseGrainingSpin} for recent reviews. While the general problem is most likely not tractable in an analytical fashion, toy models which allow for explicit computations to highlight various aspects of such a renormalisation group flow are likely to provide valuable insights. 

An example was recently provided in \cite{BWI}, where a renormalisation group flow for homogeneous and isotropic quantum states was computed using the group theoretical proposal of \cite{BodendorferCoarseGrainingAs}. Homogeneity crucially enters the construction via an ultra-local form of the dynamics, i.e. spatial points decouple and one effectively considers the sum of $N$ identical non-interacting quantum systems. The key result of \cite{BWI} was that the coarse grained dynamics of a quantum system with many small quantum numbers (spins in the loop quantum gravity language) does not agree with that of few large quantum numbers (which is captured by so called effective equations), unless the Hamiltonian operator is properly renormalised. This has important consequences for phenomenological applications of loop quantum cosmology, which are almost exclusively done in the regime of large quantum numbers with a non-renormalised Hamiltonian or using effective equations. 

While an underlying $\al{su}$(1,1) Lie algebra and SU$(1,1)$ Lie group structure entered several steps of the computations in \cite{BodendorferCoarseGrainingAs}, the deeper underlying reason for why the computations performed there (maybe somewhat surprisingly) worked out so well was not understood. In this note, we will show in section \ref{sec:GI} that the main result of \cite{BodendorferCoarseGrainingAs} follows directly from the representation theory of SU(1,1). A generalisation of the coarse graining map to arbitrary polynomials in the algebra generators and non-diagonal coherent state matrix elements is immediately implied. We verify this by an explicit computation along the lines of \cite{BodendorferCoarseGrainingAs} using Perelomov coherent states \cite{PerelomovCoherentStatesFor, PerelomovBook} in the appendix.

\section{Group interpretation of coarse graining operation} \label{sec:GI}

\subsection{Coarse physics is a captured by a representation}

We consider the $\al{su}(1,1)$ Lie algebra in a representation of the discrete class with positive magnetic quantum numbers. These representations are infinite dimensional, unitary, irreducible, and labelled by a positive half-integer\footnote{The Lie algebra $\al{su}(1,1)$ admits such representations also with real labels $j>0$, see e.g. \cite{SchliemannCoherentStatesOf}. These however generally do not exponentiate to representations of SU$(1,1)$ and we do not consider them here.} $j \in \mathbb N / 2$. In addition to the representation label $j$, basis states of the representation space are labelled by a non-negative integer $\mu \in  \mathbb N_0$, where $m = j + \mu$ is the analogue of the magnetic quantum number of SU$(2)$. We write them briefly as $ \ket{j,\mu}$, and note that the scalar product reads $\braket{j, \mu}{j', \mu'} = \delta_{j, j'}\delta_{\mu,\mu'}$ in bra-ket-notation. For more details on the representation theory of $\al{su}(1,1)$, we refer the reader to the textbook \cite{RamondBook}, as well as to the pedagogical introduction \cite{NovaesSomeBasicsOf} and to \cite{SchliemannCoherentStatesOf} for a study of several types of coherent states. 

The Lie algebra is spanned (over $\mathbb R$) by the three generators $\jz, \kx, \ky$, the last two of which can be conveniently assembled to the ladder operators $\kpm = \kx \pm i \ky$. Their action on the basis states reads 
	\begin{align}
		\jz\ket{j,\mu}&=\left(j+\mu\right)\ket{j,\mu}\\
		\kp\ket{j,\mu}&=\sqrt{(\mu+2j)(\mu+1)}\ket{j,\mu+1}\\
		\km\ket{j,\mu}&=\sqrt{\mu(\mu+2j-1)}\ket{j,\mu-1},
	\end{align}
and the Lie algebra relations follow as
\begin{eqnarray}
	\left[ \jz, \kpm \right] = \pm \kpm, ~~~~\left[ \kp, \km \right] = -2 \jz. \label{eq:Algebra}
\end{eqnarray}
The quadratic Casimir operator reads $\jz^2 - \kx^2-\ky^2$ and evaluates to $j(j-1)$ on a representation with label $j$.

Such a group structure may arise along two different routes in a quantum system. One possibility is to start as in \cite{BodendorferCoarseGrainingAs} with a classical Poisson algebra that leads to \eqref{eq:Algebra} via the quantization rule $[\cdot,\cdot] = i \{\cdot,\cdot\}$. Another possibility is to identify operators in an existing quantum theory with $\jz, \kx, \ky$ so that \eqref{eq:Algebra} follows via commutators. This second approach is more interesting in practice, especially if the various representations with label $j$ arise as a series of nested sub Hilbert spaces of the initial Hilbert space. In \cite{BWI}, this structure was key to visualise how the coarse graining operation works and to highlight the dependence of the Hamiltonian on the renormalisation scale, which turns out to be $j$. 

 Let us now describe the coarse graining operation in detail. We consider (as above) a representation with label $j$, but now take $N$ independent copies of it. A single representation space is identified as a subset of the physical system under study, and the collection of the $N$ copies as the complete system. The physical quantities corresponding to the operators $\jz, \kx, \ky$, or equivalently $\jz, \kpm$ are taken to be extensive, i.e. the corresponding properties of the complete system scale linearly with the system size, and thus linearly with $N$ when keeping the physics in each cell identical. The coarse grained physical quantities of the complete system are given by the sums 
\begin{equation}
		\KM\coloneqq\left(\prescript{}{1}{\km}+\prescript{}{2}{\km}+\ldots+\prescript{}{N}{\km}\right), \label{eq:DefCoarseKM}
	\end{equation}
and similar for $\KP$ and $\JZ$. Here, the prescript $\in 1, \ldots, N$ indicates the copy of the representation space on which a given operator acts. 

As an example, $\jz$ measures the volume of a spatial slice in homogeneous and isotropic cosmology in \cite{BodendorferCoarseGrainingAs, BWI}. Hence, one can consider the complete universe to be subdivided into $N$ non-interacting cells, each contributing the same volume. To increase the total volume, one can either increase the volume in each cell (few large quantum numbers), or increase the number of cells $N$ (many small quantum numbers). These descriptions are connected by a renormalisation group flow, which was studied in \cite{BWI}.

Due to the independence of the $N$ representation spaces and thus the commutativity of generators associated to different copies, the algebra of the coarse grained quantities \eqref{eq:DefCoarseKM} reads 
\begin{eqnarray}
	\left[ \JZ, \KPM \right] = \pm \KPM, ~~~~\left[ \KP, \KM \right] = -2 \JZ. \label{eq:AlgebraCoarse}
\end{eqnarray}
Since \eqref{eq:AlgebraCoarse} is identical to \eqref{eq:Algebra}, $\JZ$, $\KPM$ must also constitute a representation of the Lie algebra $\al{su}(1,1)$. Denoting by $\mathcal D_j$ the above representation space with label $j$, the recoupling theory for $\al{su}(1,1)$ implies (see \cite{GerryOnTheClebsch} and references therein)
\begin{equation}
	\mathcal D_{j_1} \otimes \mathcal D_{j_2} = \sum_{j = j_1 + j_2}^{\infty} \mathcal D_j . \label{eq:DDD}
\end{equation}
We will see in the following that our choice of coherent states selects only $j = j_1 + j_2$ in \eqref{eq:DDD}, so that the coarse grained system is described with a single irreducible representation. 
In the generalisation to the product of $N$ identical representations, i.e. for $\JZ$, $\KPM$, the irreducible representation with label $N j$ is selected.

\subsection{Coherent states and coarse graining map} \label{sec:CoarseGraining}

There are two important properties of the quantum states that we use for coarse graining. First, the homogeneity assumption suggests to use the same quantum state in every copy, i.e. a total symmetrisation of the wave function over the $N$ independent representation spaces. Second, the use of Perelomov coherent states (in each copy) will select only the lowest irreducible representation in \eqref{eq:DDD} and furthermore ensure that we can transfer the dynamics between the fine and coarse levels. 

Let us first look at a single copy of $\mathcal D_{j}$. Perelomov coherent states \cite{PerelomovCoherentStatesFor, PerelomovBook} for the discrete series of SU(1,1) representations used in this paper can be defined as
\begin{equation}
		\ket{j,z}=\left(1-\abs{z}^2\right)^j\sum_{\mu=0}^{\infty}\sqrt{\frac{\Gamma(\mu+2j)}{\mu\fac\Gamma(2j)}}z^\mu\ket{j,\mu}, \label{eq:DefCoh}
\end{equation}
where $ z\in\IC$ with\footnote{The restriction $|z|<1$ follows from \eqref{eq:DefCoh} being a rewriting of $e^{\xi \KP - \bar{\xi} \KM}   \ket{j, 0} = e^{2 i \Im(\xi) \KX +2 i \Re(\xi) \KY}   \ket{j, 0} =  e^{z \KP} e^{\eta \JZ} e^{-\bar{z} \KM} \ket{j, 0}$, where $\xi \in \mathbb C$ without further restrictions. This makes it manifest that \eqref{eq:DefCoh} is obtained from $\ket{j,0}$ via a transformation of the \textit{real} Lie group SU$(1,1)$. (See \cite{PerelomovBook}, section 5.2.1 for details.)} $|z|<1$. We particularly note that $\ket{j,z=0} = \ket{j,\mu=0}$ and simply write this state as $\ket{j,0}$ as no confusion can occur. The general theory of Perelomov coherent states now tells us that all states in \eqref{eq:DefCoh} can be constructed as $T(g) \ket{j, 0}$, where $g \in \text{SU}(1,1)$ and $T$ is the representation matrix of $g$. It is also important to note that there exists a non-unitary representation of SU$(1,1)$ on spinors $(z^0, \bar{z}^1) \in \mathbb C^2$. Setting $z = \frac{z^0}{\bar{z}^1}$ in \eqref{eq:DefCoh}, one finds that (see e.g. \cite{LivineGroupTheoreticalQuantization})
\begin{equation}
	U\ket{j,z} = \ket{j, U \cdot z} ~ \forall ~ U \in \text{SU}(1,1) \text{,} \label{eq:UonCS}
\end{equation}
where the action of the group element $U$ is in the infinite-dimensional unitary representation with label $j$ on the left hand side, and in the non-unitary spinor representation on the right hand side.  

On $N$ copies of $\mathcal D_{j}$, we use the product state
	\begin{equation}
		\ket{j\otimes j\otimes\ldots,z}\coloneqq\underbrace{\ket{j,z}\otimes\ket{j,z}\otimes\ldots}_{N\text{~times}},
	\end{equation}
where the totally symmetric structure is motivated by the homogeneity assumption of the physical system. 

Given these states, we can now come back to \eqref{eq:DDD} and show that only the lowest irreducible representation is selected. For this, we compute the action of the Casimir operator in the coarse representation as 
\begin{eqnarray}
		&& \left( \JZ^2 - \KX^2-\KY^2 \right)\ket{j\otimes j\otimes\ldots,z} \label{eq:CasimirProperty} \\
		&=& \left( \JZ^2 - \frac{1}{2} \KP \KM- \frac{1}{2} \KM \KP \right)\ket{j\otimes j\otimes\ldots,z} \nonumber \\
		&=& \left( \JZ^2 - \frac{1}{2} \KP \KM- \frac{1}{2} \KM \KP \right) T(g) \ket{j\otimes j\otimes\ldots,0} \nonumber \\
		&=& T(g) \left( \JZ^2 - \frac{1}{2} \KP \KM- \frac{1}{2} \KM \KP\right)  \ket{j\otimes j\otimes\ldots,0} \nonumber \\
		&= & T(g) \left( \sum_{i=1}^N \left(\prescript{}{i}{\jz}^2-\frac{1}{2}\prescript{}{i}{\kp}\prescript{}{i}{\km}-\frac{1}{2}\prescript{}{i}{\km}\prescript{}{i}{\kp}  \right)+ \sum_{i,j=1, ~~ i \neq j}^N \left(\prescript{}{i}{\jz} \prescript{}{j}{\jz} -\frac{1}{2}\prescript{}{i}{\kp}\prescript{}{j}{\km}-\frac{1}{2}\prescript{}{i}{\km}\prescript{}{j}{\kp}  \right) \right) \nonumber \\ && ~~~~~~~~~~~ ~~~~~~~~ \times \ket{j\otimes j\otimes\ldots,0} \nonumber \\ 
		&= & T(g) \left( N j (j-1) + N (N-1) j^2 \right) \ket{j\otimes j\otimes\ldots,0} \nonumber \\ 
		&= &   N j ( N j-1)  \ket{j\otimes j\otimes\ldots,z} \text{.} \nonumber
\end{eqnarray}
We observe that the Casimir operator on the coarse representation evaluates to $N j ( N j-1)$ for all coherent states, so that the coarse grained representation is labelled by $Nj$. Hence, only the representation with the smallest label survives in the decomposition \eqref{eq:DDD}. 

So far, we observed that the coarse quantities $\JZ, \KX, \KY$ form an irreducible representation with label $Nj$ when restricted to act on Perelomov coherent states. A similar property also holds for the states themselves. From the basic property $\ket{j,z} = T(g) \ket{j, 0}$ for some $g$, it follows that 
\begin{eqnarray}
	& & \ket{j\otimes j\otimes\ldots,z} \label{eq:CoarseState}\\
	&:=& \underbrace{\ket{j,z}\otimes\ket{j,z}\otimes\ldots}_{N\text{~times}} \nonumber  \\
	&=& \underbrace{T_j(g) \ket{j,0}\otimes T_j(g) \ket{j,0}\otimes\ldots}_{N\text{~times}} \nonumber \\
	&=& T_{Nj}(g) \underbrace{\ket{j,0}\otimes\ket{j,0}\otimes\ldots}_{N\text{~times}}\nonumber \\
	&=& T_{Nj}(g) \ket{Nj,0}\nonumber \\
	&=& \ket{Nj,z}, \nonumber 
\end{eqnarray}
where the identification of $\ket{j\otimes j\otimes\ldots,0}$ with $\ket{Nj,0}$ follows from comparing the $\JZ$ eigenvalue, which is $Nj$ in both cases, and the step from the third to the fourth line uses that we only select the lowest representation in the tensor product of $N$ representations with label $j$ when acting on coherent states. The last line follows from \eqref{eq:UonCS}, which ensures that the same spinor $z$ as in the first line labels the coherent state. 

We are now in a position to gather our results and summarize them in the following coarse graining map:

\begin{center}
\def\arraystretch{2.4}
  \begin{tabular}{ l | c | c }
    
     & Fine description & Coarse description \\ \hline
    Quantum state & $\underbrace{\ket{j,z}\otimes\ket{j,z}\otimes\ldots}_{N\text{~times}} $ & $\ket{Nj, z}$ \\ [20pt] \hline
    Operators & $ \left(\sum_{a=1}^N \prescript{}{a}\km\right)^p \left(\sum_{b=1}^N \prescript{}{b}\jz \right)^q  \left(\sum_{c=1}^N \prescript{}{c}\kp\right)^r$ & $\KM^p \JZ^q \KP^r$ \\ [7pt]
    \hline
  \end{tabular}
\end{center}

Our previous discussion ensures that the action of the coarse grained operators on the coarse grained states reproduces exactly the action of the fine operators on the fine states, as far as coarse observables are concerned. In other words, coarse observables can be evaluated either on fine states or on coarse states, with identical results. In particular, 
\begin{eqnarray}
	\bra{j\otimes j\otimes\ldots,\zeta}{\KM^p\, \JZ^q\, \KP^r}\ket{j\otimes j\otimes\ldots,z} = \mel{Nj,\zeta}{\KM^p\,\JZ^q\,\KP^r}{Nj,z}, \label{eq:CGExpectation}
\end{eqnarray}
where $p, q, r \in \mathbb N_0$ and $\JZ, \KPM$ act on the left hand side via \eqref{eq:DefCoarseKM}, and on the right hand side as the generators in representation $Nj$. It is easy to see that \eqref{eq:CGExpectation} implies a similar relation also for integer powers of sums of products of generators. Following the initial strategy of \cite{BodendorferCoarseGrainingAs}, we will prove \eqref{eq:CGExpectation} in the appendix via a direct calculation. Previously, this has only been shown in the case $\zeta = z$ and for powers of single generators \cite{BodendorferCoarseGrainingAs}. 

As a last remark, we recall the observation of \cite{BodendorferCoarseGrainingAs} that if the dynamics of the system is generated by a Lie algebra element, then it commutes with the coarse graining operation due to \eqref{eq:UonCS}. Hence, dynamics can be computed at either the fine or the coarse level and yields agreement for the coarse observables. This fact was used in \cite{BWI} to derive an explicit coarse graining, or renormalisation group flow for the Hamiltonian. While the coarse graining map here just tells us to change the representation of the algebra element, the successive embedding structure of the various $\al{su}(1,1)$ representation spaces into the loop quantum cosmology Hilbert space in \cite{BWI} allowed to extract a family of operators, all acting on the loop quantum cosmology Hilbert space, that depends on the renormalisation scale $j$.

\subsection{Other Lie Groups}

In this section, we will analyse which properties of our construction were crucial for obtaining the desired coarse graining properties and how they generalise to other Lie groups. 
It is well known \cite{BarutBook} that every Lie algebra \( \mathfrak{g} \) can be written in a so-called Cartan basis, which splits the algebra in a set of multiplication operators $H_j,~j=1,\ldots,r=\text{rank\,}\mathfrak{g}$ and raising and lowering operators $E_\alpha$, where \( \alpha=(\alpha_1,\ldots,\alpha_r) \) is a vector in the \emph{root system} $R$ of the complex hull \( \mathfrak{g}_{\mathds{C}} \) of $\mathfrak{g}$. The quadratic Casimir of $\mathfrak{g}$ is given by 
\begin{equation}
	\mathfrak{C}=\sum_jH_j^2+\sum_{\alpha\in R_+}E_\alpha E_{-\alpha}+E_{-\alpha}E_{\alpha}, 
\end{equation}
where $R_+$ is the set of positive roots, i.e. \( E_{\alpha\in R_+} \) are raising operators.

The construction of Perelomov coherent states works analogously for arbitrary Lie groups by acting with group elements on a fixed vector in a representation space \cite{PerelomovBook}.
The crucial point of \eqref{eq:CasimirProperty} was to show that the coarse representation is irreducible, which makes it convenient to work at the coarse level. 
	 For the analogue of \eqref{eq:CasimirProperty} to work, is sufficient to require that
	 \begin{enumerate}
	 	\item the coherent states are built from a fixed vector \( \ket{\psi_0} \) in a single irreducible representation that is annihilated by either $E_{\alpha}$ or \( E_{-\alpha} \) for every \( \alpha\in R_+ \)\footnote{Not quite incidentally, these vectors are the ones generating the ``maximally classical'' coherent states \cite[Sec. 2.4]{PerelomovBook}, i.e. those that minimise the dispersion \( \Delta\mathfrak{C} \) of the quadratic Casimir.}, 
		\item each representation occurs only once on the right hand side in the analog of \eqref{eq:DDD},
		\item the quadratic Casimir operator determines the representation uniquely.
	\end{enumerate}
	 As a consequence, the terms involving raising and lowering operators in the second sum of line 5 of equation \eqref{eq:CasimirProperty} vanish and the Casimir operator acts diagonally. 
	  Due to these properties, the coarse grained representation is also irreducible\footnote{In principle, one can drop these requirements and work with a reducible coarse grained representation, but this makes things more cumbersome.}.
	 
	 Finally, the dynamics can be transferred between the fine and coarse descriptions again if it is generated by a Lie algebra element. This is ensured by the construction of the coherent states as group elements acting on fixed vectors and the construction of coarse grained states. This also ensures that the analog of \eqref{eq:CoarseState} works out as above.

	 Examples for Lie groups that have irreducible representations containing at least one vector \( \ket{\psi_0} \) as above, i.e. the coherent states satisfy requirement 1 (\textit{but not necessarily 2 and 3}), are
		\begin{itemize}
			\item All compact, simple Lie groups, all of which have at least one highest weight vector in each irrep (e.g. SU(2), where for every irrep $j$, possible choices are the vector \( \ket{j,j} \) and \( \ket{j,-j} \)).
			\item Non-compact, semi-simple Lie groups with discrete series of irreps. These always have a lowest weight vector (e.g. SU(1,1) and the vector \( \ket{j,0} \) for every irrep in the discrete series, as well as the singleton representations).
			\item Some nilpotent groups like e.g. the Heisenberg-Weyl group \cite{PerelomovBook}.
			\item Some solvable groups, like the oscillator group \cite{PerelomovBook}.
		\end{itemize}
	For them, one needs to check on a case-by-case basis whether requirements 2 and 3 are satisfied.

Two examples beyond SU$(1,1)$ where all requirements are satisfied are given by the group SU$(2)$ as well as the so-called simple (or most degenerate / completely symmetric / class one) representations of SO$(N)$, which naturally occur in dimension-independent analogues of loop quantum gravity \cite{BTTVIII, BTTV} (see also \cite{LongCoherentIntertwinerSolution, LongPerelomovTypeCoherent} for applications of Perelomov coherent states therein). For the latter, the analogue of \eqref{eq:DDD} reads (see \cite{GirardiKroneckerProductsFor, GirardiGeneralizedYoungTableaux} or the summary in the appendix of \cite{BTTV})
\begin{equation}
	(\lambda_1,0,...,0) \otimes (\lambda_2,0,...,0) = \sum_{k = 0}^{\lambda_2}\sum_{l=0}^{\lambda_2-k} (\lambda_1 + \lambda_2 - 2k - l,l,0,...,0) \hspace{5mm} (\lambda_2 \leq \lambda_1)\text{,}
\end{equation}
	 and the restriction to Perelomov coherent states built from vectors annihilated by the raising operator selects only the (simple) highest weight representation with $k=l=0$.

Evidently, it is then possible to extend this coarse graining operation to a large class of other groups simply by group theoretical reasoning. This is especially useful as a direct algebraic proof becomes very cumbersome for $r>1$.

\section{Conclusion}

In this note, we have outlined how the results of \cite{BodendorferCoarseGrainingAs} follow from group theoretic considerations. This rephrasing is interesting because it suggests a straight forward way to generalise these results to other groups that may encode the physics of other, in particular more complicated quantum systems. The homogeneity assumption on the quantum states that crucially enters out construction is likely to be widely applicable wherever a symmetry reduction occurs, see for example \cite{GielenCosmologyFromGroup, OritiBouncingCosmologiesFrom} for a use of similar states. 

\section*{Acknowledgments}

NB was supported by an International Junior Research Group grant of the Elite Network of Bavaria. The authors would like to thank Muxin Han and Hongguang Liu for discussions on the topic.

\appendix

\section{Coherent state matrix elements} \label{sec:CSME}

	We observe that $ \bra{j,\mu}\km=\bra{j,\mu+1}\sqrt{(\mu+2j)(\mu+1)} $. From definition \eqref{eq:DefCoh}, it is straight-forward to compute matrix elements of the general form (see e.g. \cite{NovaesSomeBasicsOf})
	\begin{align}
		& \bra{j,\zeta}{\km^p\, \jz^q\, \kp^r}\ket{j,z}	\label{matrixElements} \\
			=&(2L(\zeta))^j(2L(z))^jz^{p-r}\sum_{\mu=0}^{\infty}\frac{\Gamma(\mu+p+2j)\Gamma(\mu+p+1)}{\mu\fac\Gamma(\mu+p-r+1)\Gamma(2j)}\left(\mu+p+j\right)^q\left(\bar{\zeta} z\right)^\mu. \nonumber 
	\end{align}
	
Throughout this appendix, we will be using the multinomial theorem
	\begin{equation}\label{multinomialTheorem}
		\left(\sum_{i=1}^{N}x_i\right)^n=\partsum{n}
		\frac{n\fac}{n_1\fac\ldots n_N\fac}x_1^{n_1}\ldots x_N^{n_N},
	\end{equation}
a generalised form of Vandermonde's identity
	\begin{equation}\label{vandermonde}
		\partsum{n}n\fac\binom{m_1}{n_1}\times\ldots\times\binom{m_N}{n_N}
		=\frac{\Gamma(\sum_im_i+1)}{\Gamma(\sum_im_i+1-n)},
	\end{equation}
as well as the following two identities:
	\begin{align}
		\partsum{n}\frac{n\fac}{n_1\fac\ldots n_N\fac}
		\prod_i\frac{(m_i+n_i-1)\fac}{(m_i-1)\fac}
		&=\frac{\Gamma(\sum_im_i+n)}{\Gamma(\sum_im_i)}\label{identity1}\\
		\sum_{i=0}^{k}\binom{a+i-1}{i}\binom{b+k-i-1}{k-i}&=\binom{a+b+k-1}{k}\label{identity2}.
	\end{align}
	
	\newpage
	Let us now compute the above matrix elements in the many-small-spin system (we abbreviate $\ket{z} := \ket{j\otimes j\otimes\ldots,z}$):
	\begin{align}
		&\bra{j\otimes j\otimes\ldots,\zeta}{\KM^p\, \JZ^q\, \KP^r}\ket{j\otimes j\otimes\ldots,z} \\
		=&\mel**{\zeta}{\left(\sum_i\prescript{}{i}{\km}\right)^p
			\left(\sum_i\prescript{}{i}{\jz}\right)^q
			\left(\sum_i\prescript{}{i}{\kp}\right)^r}{z}\\
		\overset{\eqref{multinomialTheorem}}{=}&\bra{\zeta}
		{
			\left(\partsum{p}\frac{p\fac}{\prod_ip_i\fac}\prod_i\prescript{}{i}{\km}^{p_i}\right)
			\left(\partsum{q}\frac{q\fac}{\prod_iq_i\fac}\prod_i\prescript{}{i}{\km}^{q_i}\right) \left(\partsum{r}\frac{r\fac}{\prod_ir_i\fac}\prod_i\prescript{}{i}{\km}^{r_i}\right)
		}
		\ket{z}\\
		=&\partsum{p}\partsum{q}\partsum{r}
		\frac{p\fac q\fac r\fac}{\prod_i p_i\fac q_i\fac r_i\fac}
		\prod_i\mel**{j,\zeta}{\km^{p_i}\,\jz^{q_i}\,\kp^{r_i}}{j,z}\\
		\overset{\eqref{matrixElements}}{=}&\partsum{p}\partsum{q}\partsum{r}
		\frac{p\fac q\fac r\fac}{\prod_i p_i\fac q_i\fac r_i\fac} \nonumber\\
		&\times\prod_i\left\{\vphantom{\sum_{\mu=0}^{\infty}}
		(1-\abs{\zeta}^2)^j(1-\abs{z}^2)^jz^{p_i-r_i}
		\right. \nonumber\\
		&\times\left.
		\sum_{\mu_i=0}^{\infty}\frac{\Gamma(\mu_i+p_i+2j)\Gamma(\mu_i+p_i+1)}{\mu_i\fac\Gamma(\mu_i+p_i-r_i+1)\Gamma(2j)}
		(\mu_i+p_i+j)^{q_i}(\bar{\zeta}z)^{\mu_i}
		\right\}\\
		=&(1-\abs{\zeta}^2)^{Nj}(1-\abs{z}^2)^{Nj}z^{p-r}
		\partsum{p}\partsum{r}\frac{p\fac r\fac}{\prod_i p_i\fac r_i\fac}
		\partsum{q}\frac{q\fac}{\prod_i q_i\fac} \nonumber\\
		&\times\sum_{\mu_1,\ldots,\mu_N=0}^{\infty}\prod_i
		\frac{\Gamma(\mu_i+p_i+2j)\Gamma(\mu_i+p_i+1)}{\mu_i\fac\Gamma(\mu_i+p_i-r_i+1)\Gamma(2j)}
		(\mu_i+p_i+j)^{q_i}(\bar{\zeta}z)^{\mu_i}\\
		\overset{\eqref{multinomialTheorem}}{=}&
		(1-\abs{\zeta}^2)^{Nj}(1-\abs{z}^2)^{Nj} z^{p-r}
		\partsum{p}\partsum{r}\frac{p\fac r\fac}{\prod_i p_i\fac r_i\fac} \nonumber\\
		&\times\sum_{\mu_1,\ldots,\mu_N=0}^{\infty}\left\{\left(\sum_i\mu_i+p+Nj\right)^q(\bar{\zeta}z)^{\sum_i\mu_i}\right. \nonumber\\
		&\hphantom{\sum_{\mu_1,\ldots,\mu_N=0}^{\infty}}
		\times\left.\vphantom{\left(\sum_i\mu_i+p+Nj\right)^q}
		\prod_i\frac{\Gamma(\mu_i+p_i+1)\Gamma(\mu_i+p_i+2j)}{\mu_i\fac\Gamma(\mu_i+p_i-r_i+1)\Gamma(2j)}
		\right\}.
	\end{align} 
	
	In the first step, we used the multinomial theorem \eqref{multinomialTheorem} to expand the products of the individual operators. Then, we used the linearity of the matrix elements and the product state property to collect the operators acting in each cell in the corresponding single-cell matrix elements, dropping the prescript $ i $ for readability. Next, we use the explicit form of these matrix elements \eqref{matrixElements} and switch the order of the summation over the $ \mu_i $ and the product over $ i $. As all sums converge absolutely, this is not a problem. Finally, we use the multinomial theorem once more to sum over the partitions of $ q $.
	
	Note that we can rewrite the combinatorial term after the product sign in the following way:
	\begin{equation}
		\frac{\Gamma(\mu_i+p_i+1)\Gamma(\mu_i+p_i+2j)}{\mu_i\fac\Gamma(\mu_i+p_i-r_i+1)\Gamma(2j)}
		=\frac{(\mu_i+p_i)\fac r_i\fac}{\mu_i\fac}
		\binom{\mu_i+p_i}{r_i}\binom{\mu_i+p_i+2j-1}{\mu_i+p_i}.
	\end{equation}
	
	This allows us to use Vandermonde's identity \eqref{vandermonde} to sum over the partitions of $ r $:
	\begin{align}
		\mel{\zeta}{\ldots}{z}\overset{\eqref{vandermonde}}{=}&(1-\abs{\zeta}^2)^{Nj}(1-\abs{z}^2)^{Nj} z^{p-r}
		\partsum{p}\frac{p\fac}{\prod_i p_i\fac} \nonumber\\
		&\times\sum_{\mu_1,\ldots,\mu_N=0}^{\infty}
		\left(\sum_i\mu_i+p+Nj\right)^q(\bar{\zeta}z)^{\sum_i\mu_i} \frac{\Gamma(\sum_i\mu_i+p+1)}{\Gamma(\sum_i\mu_i+p-r+1)} \nonumber\\
		&\times\prod_i\frac{(\mu_i+p_i)\fac r_i\fac}{\mu_i\fac}\binom{\mu_i+p_i+2j-1}{\mu_i+p_i}.
	\end{align}
	
	Let us again rewrite the term after the product sign:
	\begin{equation}
		\frac{(\mu_i+p_i)\fac r_i\fac}{\mu_i\fac}\binom{\mu_i+p_i+2j-1}{\mu_i+p_i}
		=\binom{\mu_i+2j-1}{\mu_i}\frac{(\mu_i+p_i+2j-1)\fac}{(\mu_i+2j-1)\fac}.
	\end{equation}
	
	Using the identity \eqref{identity1}, we can finally carry out the last remaining partition sum over $ p $:
	\begin{align}
		\mel{\zeta}{\ldots}{z}\overset{\eqref{identity1}}{=}&
		(1-\abs{\zeta}^2)^{Nj}(1-\abs{z}^2)^{Nj} z^{p-r} \nonumber\\
		&\times
		\sum_{\mu_1,\ldots,\mu_N=0}^{\infty}
		\left\{
		\left(\sum_i\mu_i+p+Nj\right)^q(\bar{\zeta}z)^{\sum_i\mu_i} \frac{\Gamma(\sum_i\mu_i+p+1)}{\Gamma(\sum_i\mu_i+p-r+1)}
		\right. \nonumber\\
		&\hphantom{	\sum_{\mu_1,\ldots,\mu_N=0}^{\infty}}\times
		\left.\vphantom{\left(\sum_i\mu_i+p+Nj\right)^q}
		\frac{\Gamma(\sum_i\mu_i+p+2Nj)}{\Gamma(\sum_i\mu_i+2Nj)}
		\prod_i\binom{\mu_i+2j-1}{\mu_i}
		\right\}\\
		=&(1-\abs{\zeta}^2)^{Nj}(1-\abs{z}^2)^{Nj} z^{p-r} \nonumber\\
		&\times\sum_{k_1=0}^{\infty}\sum_{k_2=0}^{k_1}\ldots\sum_{k_N=0}^{k_{N-1}}
		(k_1+p+Nj)^q(\bar{\zeta}z)^{k_1}
		\frac{\Gamma(k_1+p+1)\Gamma(k_1+p+2Nj)}{\Gamma(k_1+p-r+1)\Gamma(k_1+2Nj)} \nonumber\\
		&\times\binom{k_N+2j-1}{k_N}\binom{k_{N-1}-k_N+2j-1}{k_{N-1}-k_N}\times\ldots\times
		\binom{k_1-k_2+2j-1}{k_1-k_2},
	\end{align}
	where in the second step, we simply reordered the sum over the $ \mu_i $.
	
	From this form, it is apparent that we can repeatedly use the identity \eqref{identity2} to sum over all $ k_{i>1} $. Renaming $ k_1\to\mu $, we get
	\begin{align}
		\mel{\zeta}{\ldots}{z}\overset{\eqref{identity2}}{=}&
		(1-\abs{\zeta}^2)^{Nj}(1-\abs{z}^2)^{Nj} z^{p-r} \nonumber\\
		&\times
		\sum_{\mu=0}^{\infty}(\mu+p+Nj)^q(\bar{\zeta}z)^\mu\frac{\Gamma(\mu+p+1)\Gamma(\mu+p+2Nj)}{\mu\fac\Gamma(\mu+p-r+1)\Gamma(2Nj)} \nonumber\\
		=&\mel{Nj,\zeta}{\km^p\,\jz^q\,\kp^r}{Nj,z} , 
	\end{align}
	where the generators in the last line act in the representation $Nj$. Relabelling $\jz \rightarrow \JZ$ and $\kpm \rightarrow \KPM$ finally leads to \eqref{eq:CGExpectation}.


\begin{thebibliography}{10}

\bibitem{BodendorferCoarseGrainingAs}
N.~Bodendorfer and F.~Haneder, ``{Coarse graining as a representation
  change},'' {\em Phys. Lett. B} {\bf 792} (2019) 69--73, {\tt arXiv:1811.02792
  [gr-qc]}.

\bibitem{BWI}
N.~Bodendorfer and D.~Wuhrer, ``{Renormalisation with SU(1, 1) coherent states
  on the LQC Hilbert space},'' {\em Class. Quantum Gravity} (2020) {\tt
  arXiv:1904.13269 [gr-qc]}.

\bibitem{ThiemannRenormalisationReview}
T.~Thiemann, ``{Canonical Quantum Gravity, Constructive QFT and
  Renormalisation},'' {\tt arXiv:2003.13622 [gr-qc]}.

\bibitem{SteinhausCoarseGrainingSpin}
S.~Steinhaus, ``{Coarse graining spin foam quantum gravity -- a review},'' {\tt
  arXiv:2007.01315 [gr-qc]}.

\bibitem{PerelomovCoherentStatesFor}
A.~M. Perelomov, ``{Coherent states for arbitrary Lie group},'' {\em Commun.
  Math. Phys.} {\bf 26} (1972) 222--236, {\tt arXiv:math-ph/0203002}.

\bibitem{PerelomovBook}
A.~M. Perelomov, {\em {Generalized Coherent States and Their Applications}}.
\newblock Springer, Berlin, 1986.

\bibitem{SchliemannCoherentStatesOf}
J.~Schliemann, ``{Coherent states of su(1,1): correlations, fluctuations, and
  the pseudoharmonic oscillator},'' {\em J. Phys. A Math. Theor.} {\bf 49}
  (2016) 135303, {\tt arXiv:1508.04549 [quant-ph]}.

\bibitem{RamondBook}
P.~Ramond, {\em {Group Theory. A Physicist's Survey.}}
\newblock Cambridge University Press, 2010.

\bibitem{NovaesSomeBasicsOf}
M.~Novaes, ``{Some basics of su(1,1)},'' {\em Rev. Bras. Ensino Fis.} {\bf 26}
  (2004) 351--357.

\bibitem{GerryOnTheClebsch}
C.~C. Gerry, ``{On the Clebsch–Gordan problem for SU(1,1): Coupling
  nonstandard representations},'' {\em J. Math. Phys.} {\bf 45} (2004)
  1180--1190.

\bibitem{LivineGroupTheoreticalQuantization}
E.~R. Livine and M.~Martin-Benito, ``{Group theoretical quantization of
  isotropic loop cosmology},'' {\em Phys. Rev. D} {\bf 85} (2012) 124052, {\tt
  arXiv:1204.0539 [gr-qc]}.

\bibitem{BarutBook}
A.~O. Barut and R.~Raczka, {\em {Theory of group representations and
  applications}}.
\newblock PWS, Warszawa, 1977.

\bibitem{BTTVIII}
N.~Bodendorfer, T.~Thiemann, and A.~Thurn, ``{Towards Loop Quantum Supergravity
  (LQSG)},'' {\em Phys. Lett. B} {\bf 711} (2012) 205--211, {\tt
  arXiv:1106.1103 [gr-qc]}.

\bibitem{BTTV}
N.~Bodendorfer, T.~Thiemann, and A.~Thurn, ``{On the implementation of the
  canonical quantum simplicity constraint},'' {\em Class. Quantum Gravity} {\bf
  30} (2013) 045005, {\tt arXiv:1105.3708 [gr-qc]}.

\bibitem{LongCoherentIntertwinerSolution}
G.~Long, C.-Y. Lin, and Y.~Ma, ``{Coherent intertwiner solution of simplicity
  constraint in all dimensional loop quantum gravity},'' {\em Phys. Rev. D}
  {\bf 100} (2019) 064065, {\tt arXiv:1906.06534 [gr-qc]}.

\bibitem{LongPerelomovTypeCoherent}
G.~Long and N.~Bodendorfer, ``{Perelomov type coherent states of SO(D + 1) in
  all dimensional loop quantum gravity},'' {\tt arXiv:2006.13122 [gr-qc]}.

\bibitem{GirardiKroneckerProductsFor}
G.~Girardi, A.~Sciarrino, and P.~Sorba, ``{Kronecker products for SO(2p)
  representations},'' {\em J. Phys. A. Math. Gen.} {\bf 15} (1982) 1119--1129.

\bibitem{GirardiGeneralizedYoungTableaux}
G.~Girardi, A.~Sciarrino, and P.~Sorba, ``{Generalized Young tableaux and
  Kronecker products of SO(n) representations},'' {\em Phys. A Stat. Mech. its
  Appl.} {\bf 114} (1982) 365--369.

\bibitem{GielenCosmologyFromGroup}
S.~Gielen, D.~Oriti, and L.~Sindoni, ``{Cosmology from Group Field Theory
  Formalism for Quantum Gravity},'' {\em Phys. Rev. Lett.} {\bf 111} (2013)
  031301, {\tt arXiv:1303.3576 [gr-qc]}.

\bibitem{OritiBouncingCosmologiesFrom}
D.~Oriti, L.~Sindoni, and E.~Wilson-Ewing, ``{Bouncing cosmologies from quantum
  gravity condensates},'' {\em Class. Quantum Gravity} {\bf 34} (2017) 04LT01,
  {\tt arXiv:1602.08271 [gr-qc]}.

\end{thebibliography}
		
\end{document}